\documentclass[letterpaper,useAMS,usenatbib]{mn2e}
\usepackage{aas_macros}
\usepackage{times}
\usepackage{amssymb}
\usepackage{graphicx}
\usepackage{rotating}
\usepackage{longtable}
\usepackage{url}

\newcommand{\ltsim}{\raisebox{-1mm}{$\stackrel{<}{\sim}$}}

\begin{document}

\title[Low-power radio galaxy environments at $z\sim0.5$ in the
SXDF]{Low-power radio galaxy environments in the {\it Subaru/XMM-Newton Deep
Field} at $\mathbf{z\sim0.5}$ }

\author[J. E. Geach et al.]{J.\ E.\ Geach$^1$\thanks{E-mail: j.e.geach@durham.ac.uk},
C.\ Simpson$^{2}$, S.\ Rawlings$^3$, A.\ M.\ Read$^4$ \& M.\ Watson$^4$\\
\\
\noindent $^1$Department of Physics, Durham University, 
South Road, Durham. DH1 3LE. UK.\\
\noindent $^2$Astrophysics Research Institute, Liverpool John Moores
University, Twelve Quays House, Egerton Wharf, Birkenhead, CH41 1LD. UK.\\
\noindent $^3$Astrophysics, Department of Physics, Denys Wilkinson Building, 
Keble Road, Oxford. OX1 3RH. UK.\\
\noindent $^4$Department of Physics and Astronomy, University of Leicester, 
Leicester, LE1 7RH. UK.}


\date{}

\pagerange{\pageref{firstpage}--\pageref{lastpage}} \pubyear{2007}

\maketitle

\label{firstpage}

\begin{abstract}
We present multi-object spectroscopy of galaxies in the immediate
(Mpc-scale) environments of four low-power ($L_{\rm 1.4 GHz} \lesssim 
10^{25}$\,W\,Hz$^{-1}$) radio galaxies at $z\sim0.5$, selected from the 
{\it Subaru/XMM-Newton Deep Field}. 
We use the spectra to calculate velocity dispersions and central
redshifts of the  groups the radio galaxies inhabit, and combined with 
{\it XMM-Newton} (0.3--10\,keV) X-ray observations investigate the 
$L_X$--$\sigma_v$ and $T_X$--$\sigma_v$ scaling relationships. All the radio
galaxies reside in moderately rich groups -- intermediate environments
between poor groups and rich clusters, with remarkably similar X-ray
properties. We concentrate our discussion on our best statistical
example that we interpret as a low-power (FR\,I) source triggered within a sub-group,
which in turn is interacting with a nearby group of galaxies,
containing the bulk of the X-ray emission for the system -- a basic scenario
which can be compared to more powerful radio sources at both high ($z>4$) and
low ($z<0.1$) redshifts. This suggests that galaxy-galaxy interactions
triggered by group mergers may play an
important role in the life-cycle of radio galaxies at all epochs and
luminosities. 
\end{abstract}


\section{Introduction}
\label{sec:intro}

\begin{table*}
\caption{Summary of radio galaxy targets. I.D.'s, positions and fluxes
  are taken from an early version of the 1.4\,GHz Very Large Array
  radio mosaic presented in Simpson et al.\ 2006 (S06). In this paper for
convenience we shortern IAU identifications of the radio galaxies to
JEG\,1--4. We quote radio luminosities here for completeness, however
we discuss the redshift determination in \S3.2.}
\begin{tabular}{lllccccc}
\hline
I.D. & IAU I.D. & S06 I.D.  & R.A. & Dec. &  $S^{\rm SXDF}_{1.4\rm\,GHz}$ &
$L_{1.4\rm\,GHz}$ \cr
 &  & &  (J2000.0) & (J2000.0) &  (mJy) & ($10^{24}$\,W\,Hz$^{-1}$) \cr
 \hline
 JEG\,1 & VLA\,J021945$-$04535 & $-$  & 02\ 19\ 45.3 & $-$04\ 53\ 33 &
 11.86$\pm$0.13 & 4.36$\pm$0.05 \cr
 JEG\,2 & VLA\,J021823$-$05250 & VLA\,0011 & 02\ 18\ 23.5 & $-$05\ 25\ 00 &
 7.95$\pm$0.10 & 14.35$\pm$0.18 \cr
 JEG\,3 & VLA\,J021737$-$05134 & VLA\,0033 & 02\ 17\ 37.2 & $-$05\ 13\ 28 &
 2.37$\pm$0.06 & 4.26$\pm$0.11 \cr
 JEG\,4 & VLA\,J021842$-$05328 & VLA\,0065  & 02\ 18\ 42.1 & $-$05\ 32\ 51
 &
 0.96$\pm$0.08 & 0.48$\pm$0.04 \cr
 \hline
 \end{tabular}
 \end{table*}

\begin{figure*}
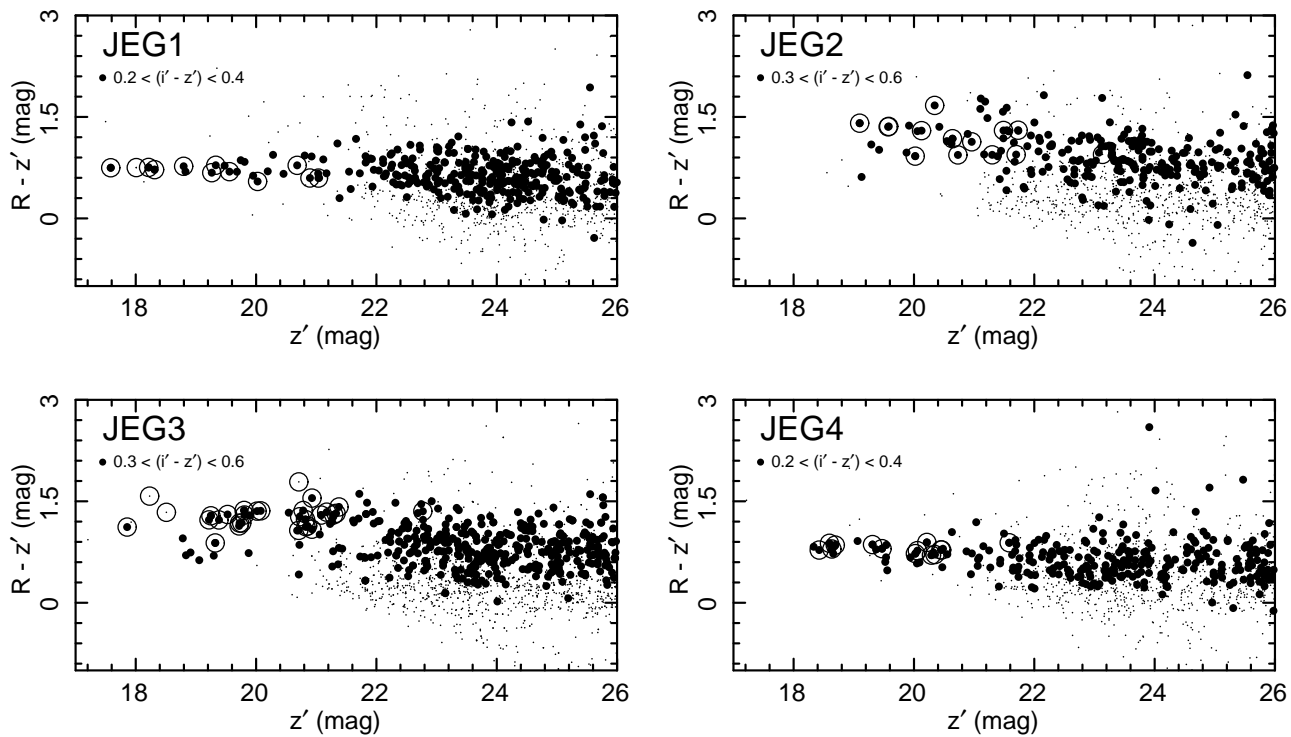

\begin{center}
\begin{turn}{-90}\includegraphics[width=2in]{figure1a.ps}\end{turn}\,\,\,\begin{turn}{-90}\includegraphics[width=2in]{figure1b.ps}\end{turn}\\
\begin{turn}{-90}\includegraphics[width=2in]{figure1c.ps}\end{turn}\,\,\,\begin{turn}{-90}\includegraphics[width=2in]{figure1d.ps}\end{turn}
\end{center}
\caption{$R-z'$ vs. $z'$ Field-corrected colour-magnitude diagrams for each
mask. Photometry was extracted from the {\it SXDF} catalog by choosing all
galaxies within an aperture of radius 2.5$'$ centered on the brightest
target galaxy. 
Colours are measured in 2$''$ apertures, and we also indicate a secondary colour,
$i'-z'$ (arbitrarily chosen), to emphasise the location of the red-sequence
as larger points. The colours of targeted galaxies in this work are
indicated by larger open circles.}
\end{figure*}

In recent years, feedback mechanisms from radio galaxies (that is, galaxies
which host radio-loud active galactic nuclei) have become the method of
choice for curtailing the bright end of the galaxy luminosity function
in models of galaxy formation, since models
without feedback tend to over-produce the number of very luminous
galaxies compared to what is observed (e.g. Bower et al. 2006). Many
high-redshift radio galaxies lie in
(proto-)cluster environments (e.g. Venemans et al. 2003) and the energies 
provided by their radio jets (the
bulk kinetic power of the radio jets is several orders of magnitude
larger than the radio luminosity; e.g., Rawlings \& Saunders 1991) is
sufficient to shut down star forming activity throughout a cluster (Rawlings \&
Jarvis 2004). Furthermore, it has been suggested that powerful radio
sources may be triggered by galaxy--galaxy interactions during the
merging of subcluster units (Ellingson, Yee \& Green 1991; Simpson \&
Rawlings 2002).  

Lower luminosity sources are also believed to be vital in
shaping the galaxy luminosity function by providing a low power, but
high duty cycle, mode of feedback (e.g., Croton et al.\ 2005; Bower et
al.\ 2006; Best et al.\ 2006). This mode of feedback appears to be
most common in the most massive galaxies, which are, of course,
preferentially located in the centres of galaxy clusters. Therefore
there appears to be significant interplay between the formation and
evolution of both the radio sources and their cluster environments.
Studying the environments of low-power radio sources is likely
a promising route towards understanding radio galaxy feedback. However,
while the radio galaxy itself can influence the thermodynamic history
of its environment (i.e. by depositing energy), the picture is
complicated by global environmental processes such as the merging of
sub-cluster units, which can temporarily boost the X-ray
luminosity and temperature of the intracluster medium (e.g. Randall,
Sarazin \& Ricker\ 2002), making such systems more readily detectable
in X-ray surveys (a matter of caution in using such X-ray detected systems as Cosmological 
probes). It is essential to perform detailed
studies of the radio, optical, and X-ray properties of putative dense
regions in the cosmic web to ascertain how mergers and radio source
activity affect the life-cycle of clusters.

The luminosity and temperature of a cluster are known to follow a scaling relationship
(the $L_X$--$T_X$ relation), $L_X \propto T_X^\alpha$,  where $\alpha$
has been measured in the range $\sim$2.7--3 (Edge \& Stewart 1991; David, Jones \&
Forman 1995; Allen \& Fabian 1998; Markevitch 1998; Arnaud \& Evrard 1999)
This is at odds with that expected
for clusters formed by gravitational structure formation, with $L_X
\propto T_X^2$ (Kaiser 1986). It is thought that the disparity is due
to the interruption of
cooling by feedback from active galaxies, which can impart energy via
outflow shocks (for example from radio-jets or super-winds), or by
performing work by sub-sonically inflating bubbles in the intracluster
medium (ICM; Sijacki \& Springel 2006). Best\ et~al.\ (2005) suggest
that, given the short life-times of radio-loud AGN
(10--100\,Myr), episodic triggering must occur in order for feedback to
quench cooling. However, much lower luminosity sources ($L_{\rm
1.4\,GHz} < 10^{25}$\,W\,Hz$^{-1}$) can impart energy into the IGM
with a higher duty cycle than the more powerful, and rarer, sources
(Best~et~al.\ 2006). 

McLure et al.\ (2004) have recently constructed a
sample of radio galaxies at $z\sim0.5$ which spans three orders of
magnitude in radio luminosity, but even the faintest of these
sources are close to the Fanaroff--Riley break (Fanaroff \& Riley 1974) and it
is desirable to push to even fainter radio luminosities to fully
investigate the radio galaxy--cluster symbiosis. In this paper we present
multi-object spectroscopy of galaxies around four low-power ($L_{\rm
1.4\,GHz} <10^{25}$\,W\,Hz$^{-1}$) radio galaxies at $z\sim0.5$ in the {\it
Subaru/XMM-Newton Deep Field (SXDF)} (Sekiguchi et al.\ 2001, 2006 in prep). 
The resultant spectroscopic redshifts
allow us to investigate the environments of these radio sources which
could provide a large proportion of the local volume averaged heating
rate if the AGN luminosity function is flat below $L_{\rm 1.4\,GHz}
\sim10^{25}$\,W\,Hz$^{-1}$ (Best~et~al.~2006). We combine these data
with X-ray observations of the radio galaxies' environments in order to
investigate the global environmental properties -- namely the
$L_X-\sigma_v$ and $T_X-\sigma_v$ relations. This parameter space can
be used as a diagnostic of the thermodynamic history of groups or
clusters, since if an AGN is significantly feeding energy back into
the intracluster medium (ICM), then the group or cluster might be
expected to deviate from the standard empirical
$L_X(T_X)-\sigma_v$ relation for massive systems. 
Throughout we adopt a flat geometry with $\Omega_{\rm m}=0.3$,
$\Omega_{\Lambda}=0.7$ and $H_0 = 100h$ km\,s$^{-1}$\,Mpc$^{-1}$ where $h =
0.75$. All magnitudes in this paper are on the AB scale.

\begin{figure}
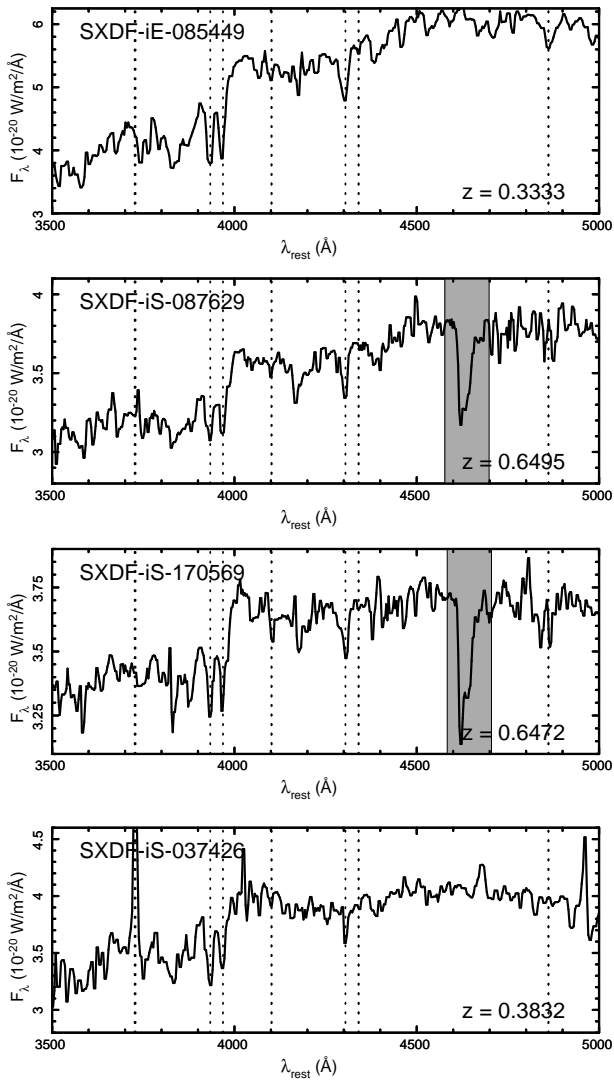

\includegraphics[width=3.25in]{figure2a.ps}
\includegraphics[width=3.25in]{figure2b.ps}
\includegraphics[width=3.25in]{figure2c.ps}
\includegraphics[width=3.25in]{figure2d.ps}
\caption{Successfully cross-correlated spectra for the radio galaxies
field of (top-down) JEG\,1--4. The spectra are plotted in the
rest-frame. As a guide, we identify as dotted vertical lines the spectral
features {\sc [Oii]}, K, H, H$\delta$, G, H$\gamma$, and H$\beta$. We do not
plot wavelengths blueward of 3500\AA\ or redward of 5000\AA\ (restframe)
since in these regimes the response and signal-to-noise of the spectra do not
allow accurate analysis. Note that the shaded region highlights a sky
absorption feature.}
\end{figure}

\section{Observations \& Reduction}

\subsection{Sample selection and spectroscopy}

Targets were selected from an early version of            
the 1.4-GHz Very Large Array radio mosaic presented in Simpson et al.\           
(2006), and listed in Table~1. The limited red sensitivity of the
Low Dispersion Survey Spectrograph (LDSS-2, Allington-Smith\ et\ al. 1994) 
placed an upper limit of $z\sim0.7$ as appropriate for our study, so we
selected radio sources whose optical counterparts are resolved and have
$19<R<20.5$,
since the tightness of the Hubble diagram for radio galaxies allows              
redshifts to be estimated from a single magnitude, and we aim to target
moderate redshift environments (Cruz et al. 2007). Colour--magnitude
diagrams were produced for $2.5'$ regions around each radio
galaxy and the four optically brightest radio galaxies whose
environments displayed a red sequence were selected for spectroscopic
follow-up. This selection was simply performed `by eye' -- if the
colour-magnitude diagram displayed an overdensity of points at
$18<m_{\rm RG}<20.5$ with tight scatter in colour space.
Note that local X-ray properties were not taken into account when
choosing the target radio galaxies -- this is simply an optical
selection of radio galaxies.

Masks were designed using the LDSS-2 mask preparation
software,
with objects along the red sequence (typically using a colour cut of width
0.3\,magnitudes) included in the input file (Figure 1). This software allows                
relative priorities to be assigned to targets, which were set to be              
equal to the magnitudes of the galaxies, to increase the likelihood of           
successfully determining redshift.

Observations were conducted using the medium-red (300 lines/mm,
$\lambda_{0} = 5550$\AA) grism of LDSS-2 in multi-object mode on the 6.5-m 
Magellan telescope in Las Campanas, Chile over the nights 17--20 October
2003 (effective resolution 13.3\AA). 
The MOS frames were trimmed for overscan, flat-fielded, rectified, cleaned
and wavelength calibrated using a sequence of Python routines (D. Kelson,
priv. comm.); sky-subtraction and spectrum extraction was performed in
IDL. Due to a  lack of standard-star observations, for flux-calibration and
to correct for the wavelength dependent response of the detector, we
extracted  {\it BVRi$'$z$'$} photometry for the targets from the {\it
 Subaru/XMM-Newton  Deep Field (SXDF)} catalog. Using the five
sources in each mask with the strongest signal-to-noise ratio (SNR), we fit
mask-specific response curves to the 1-D spectra, which are then applied to
each  extracted target spectrum. Although flux-calibration and response
correction is not necessary for redshift determination via
cross-correlation, we perform these steps in order to present a sample
of the spectra in Figure 2.

\subsection{X-ray data}

The {\it SXDF} incorporates a deep, large-area X-ray mosaic
with XMM-Newton, consisting of 7
overlapping pointings covering
$\sim$1.3\,deg$^{2}$ region of the high-Galactic latitude sky with an exposure
time of 100\,ks in the central field (in separate exposures) and
50\,ks in the flanking fields (see Table~2). Four of the pointings
were carried out in August 2000, and the remaining three were made in
August 2002 and January 2003.

\
\begin{table}
\caption[]{Overview of the {\it XMM-Newton} observations. Tabulated are the
Field IDs, the centres of the pointings and the exposure times of the relevant
high-background-cleaned datasets for each camera: MOS\,1,
MOS\,2 and PN. Note that the coordinates are in J\,2000.}
\begin{tabular}{lccc}
\noalign{\smallskip}
\hline
Field  & R.A.       & Dec.       &  Exp. time [MOS\,1~~MOS\,2~~PN]   \\ 
       & hh~~mm~~ss & $\circ$~~$'$~~$''$            & (ksec)        \\ \hline
SDS-1a & 02\ 18\ 00 & $-$05\ 00\ 00   & $-$~~$-$~~40   \\
SDS-1b & 02\ 18\ 00 & $-$05\ 00\ 00   & $-$~~$-$~~42   \\
SDS-2  & 02\ 19\ 36 & $-$05\ 00\ 00   & 47~~48~~40 \\
SDS-6  & 02\ 17\ 12 & $-$05\ 20\ 47   & 49~~49~~47 \\
SDS-7  & 02\ 18\ 48 & $-$05\ 20\ 47   & 40~~41~~35 \\
\noalign{\smallskip}
\hline
\end{tabular}
\end{table}

All the {\it XMM-Newton} EPIC data, i.e. the data from the two MOS cameras
and the single PN camera, were taken in full-frame mode with the thin
filter in place. These data have been
reprocessed using the standard procedures in the most recent
{\it XMM-Newton} SAS (Science Analysis System) $-$ v.6.5. All the datasets
where at least one of the radio galaxies was seen to fall within the
field-of-view (FOV) of the detectors were then analysed
further. JEG\,3 was observed in pointings SDS-1a (PN),
SDS-1b (PN) and SDS-6 (M1, M2, PN) (part   of the field lies off the edge of the
MOS FOVs in SDS1a and SDS1b). Both JEG\,2 and JEG\,4 were
observed in pointing SDS-7 (MOS\,1, MOS\,2, PN). Finally JEG\,1 was
observed in pointing SDS-2 (MOS\,1, MOS\,2, PN). Periods of high-background
were filtered out of each dataset by creating a high-energy
10$-$15\,keV lightcurve of single events over the entire field of
view, and selecting times when this lightcurve peaked above 0.75\,ct
s$^{-1}$ (for PN) or 0.25\,ct s$^{-1}$ (for MOS). It is known that
soft proton flares can affect the background at low energies at
different levels and times than high energies (e.g. Nevalainen et al.\
2005). This was not seen to be
the case in the datasets analysed here however, as full energy-band
(1--10\,keV) lightcurves (extracted from off-source regions) showed
essentially identical times of flaring activity as the high energy
lightcurves. The high-background times in each dataset are seen to be
very distinct and essentially unchanging with energy-band. The exposure times
of the resulting cleaned relevant datasets are given in Table~2.

Source spectra were extracted from the relevant datasets from apertures
centred on the source positions. Extraction radii, estimated from
where the radial surface brightness profiles were seen to fall to the
general surrounding background level, were set to 50$''$, 21$''$, 36$''$ and
47$''$ for JEG\,1--4 respectively. Note that the surface brightness
profiles were co-added for all 3 EPIC cameras for improved S/N. 
Standard SAS source-detection algorithms,
uncovered a large number of X-ray sources, both
point-like and extended. Data due
to contaminating sources lying within the extraction regions
needed to be removed from the spectra. This proved necessary
only for JEG\,3, where two sources were detected just within
the extraction region. Data from the two sources were removed
to a radius of 28$''$ (and the extraction area is thus reduced by
$\sim10$\%). Background spectra were extracted from each cleaned
dataset from an annulus of inner radius 60$''$ and outer radius
180$''$ around each position. As point sources were seen to
contaminate these larger-area background spectra, the data from these
detected sources were removed from the background spectra to a radius
of 60$''$. We have verified that all the X-ray sources are clearly extended 
using both simple (sliding box) and more sophisticated
wavelet analysis techniques. The possibility that JEG 2 is an
unresolved source complex cannot of course be completely ruled out
given its low signal to noise. Given the relative weakness of all of
the X-ray sources, we consider that the extraction radii used are close
to optimum to provide the best signal to noise in the extracted spectra.
Ancillary Response Files (ARFs) containing the telescope
effective area, filter transmission and quantum efficiency curves were created 
for the cluster spectra (taking into account the fact that the
extraction regions are extended), and were checked
to confirm that the correct extraction area calculations (complicated
with the exclusion of contaminating point sources, etc.) had been
performed. 
Finally Redistribution Matrix Files (RMFs) files were generated. These files
describe, for a given incident X-ray photon energy, the observed photon energy
distribution over the instrument channels. These can be used in
conjunction with the ARFs to perform X-ray spectral analysis -- we use the most
commonly used analysis package: {\sc xspec}. 
The source spectral channels were
binned together to give a minimum of 20 net counts per bin. A detailed
discussion of the X-ray spectral fitting and the results is given in
\S3.4

\begin{center}
\begin{figure}
\includegraphics[width=3.25in]{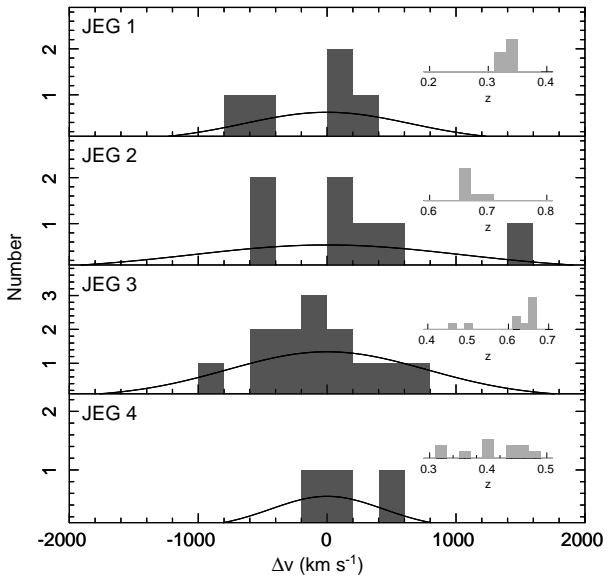}
\caption{Relative velocity distribution of galaxies where successful
redshift-determination was performed. Relative velocities are in the
cluster frame, over a range $\pm2000$\,km\,s$^{-1}$. 
Insets show the distribution in
redshift space, over a wider velocity range.
We overlay Gaussian profiles based on the fitted
velocity dispersion in each case.}
\end{figure}
\end{center}

\section{Analysis}

\subsection{Colour-magnitude diagrams}

In Figure~1 we plot the $R-z'$ vs. $z'$ colour-magnitude relations (CMRs)
for each field, which have been constructed from the photometry of all
galaxies within 2.5$'$ of each radio-galaxy  target. 
We  correct the colour-magnitude diagrams for field contamination using a
method similar to Kodama et al. (2004) and Kodama \& Bower (2001), which we
will briefly describe here. First,  colour-magnitude space is divided into
a coarse grid ($\Delta z' = 2$, $\Delta(R-z') = 0.8 $), and the number of
galaxies within each grid-square counted: $N^{\rm gal}_{ij}$.  An identical
count is performed for a control field, $N^{\rm control}_{ij}$, where
control galaxies are selected within an aperture of identical radius to the
one used for each target mask, but centered on a random point on the sky
(but within the catalog). We then assign the $ij^{\rm th}$ colour-magnitude
bin the probability that the galaxies within it are field 
members: $P_{ij} = N^{\rm control}_{ij}/(N^{\rm gal}_{ij}+N^{\rm
  control}_{ij})$. For each galaxy in colour-magnitude space, we pick a
random number, $p$,  in the interval \{0,1\} and compare it to $P_{ij}$; a
galaxy is removed from the diagram if $p < P_{ij}$. In order to reduce
effects of cosmic-variance (e.g. randomly picking another cluster or void
as the control sample), we build up an average probability map by repeating
the control sample count and probability calculation 10 times 
around different random positions in the catalog.  

There are clear red-sequences around all four radio galaxies, and to
emphasise this, we also plot a secondary colour $i-z'$, and the locations of
galaxies in colour space targeted for spectroscopy. The secondary colour-cut is chosen
merely to highlight the red-sequence.

\begin{center}
\begin{figure*}
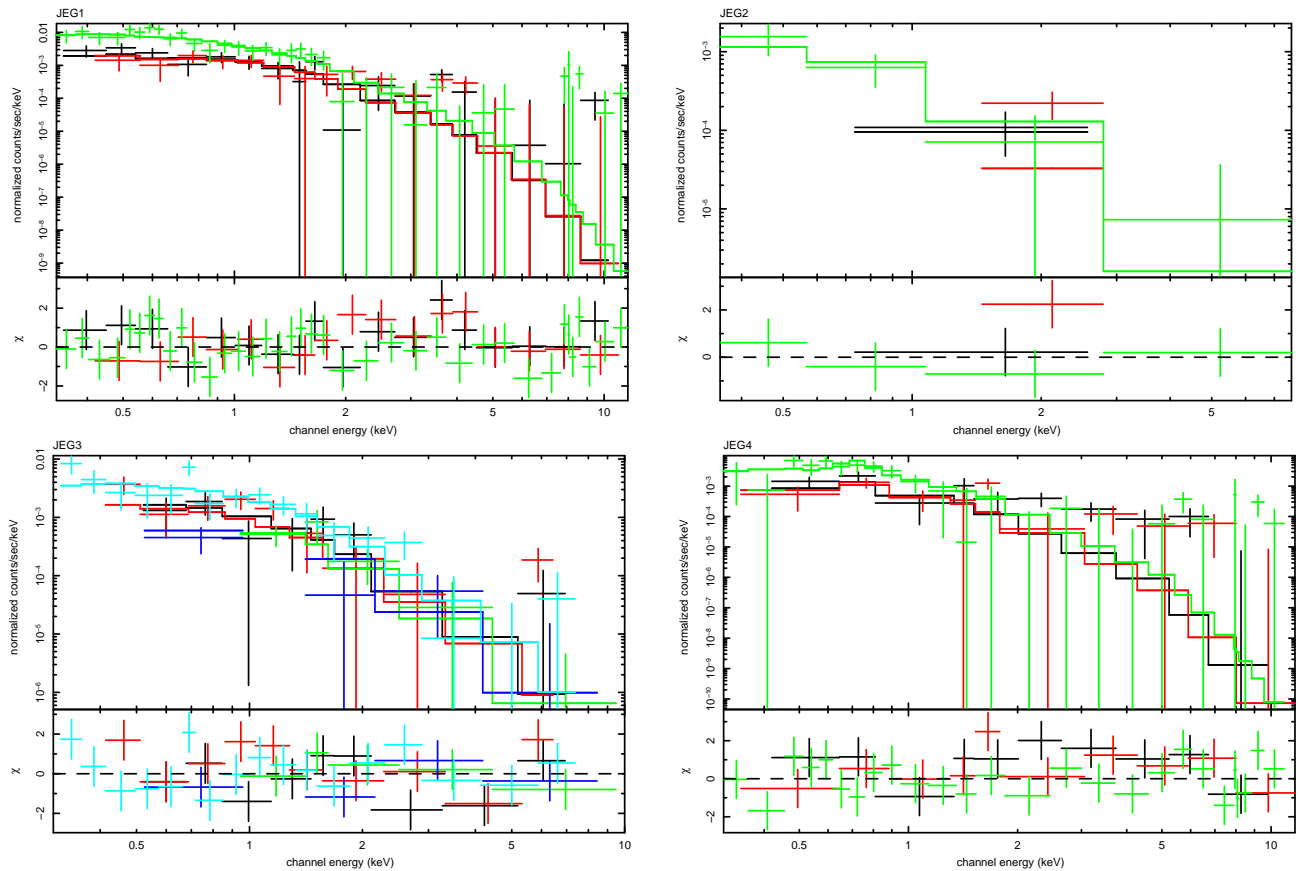

\begin{turn}{-90}
\includegraphics[width=2.25in]{figure4a.ps}
\end{turn}\hspace{5mm}
\begin{turn}{-90}
\includegraphics[width=2.25in]{figure4b.ps} 
\end{turn}
\vspace{5mm}
\begin{turn}{-90}
\includegraphics[width=2.25in]{figure4c.ps}
\end{turn}\hspace{5mm}
\begin{turn}{-90}
\includegraphics[width=2.25in]{figure4d.ps} 
\end{turn}
\caption{The {\it XMM-Newton} EPIC spectra of (row-wise from top left) JEG\,1--4, 
together with the best-fitting thermal plasma model fitted using {\sc xspec}. 
$\chi^{2}$ differences are shown in the lower panels of each plot. The
colour-codings correspond to the MOS\,1, MOS\,2 and PN cameras and field
pointings (see Table\,2 and \S2.2). For JEG\,1 (SDS-2), JEG\,2 (SDS-7)
\& JEG\,4 (SDS-7) black is MOS\,1, red is MOS\,2 and green is PN. 
For JEG\,3 black is PN (SDS-1a), red is PN (SDS-1b), green is MOS\,1 (SDS-6), 
blue is MOS\,2 (SDS-6) and cyan is PS (SDS-6). The resulting cluster
parameters of $L_X$ and $T_X$ are presented in Table~4.}
\end{figure*}
\end{center}

\subsection{Redshift determination and velocity dispersions}

For redshift determination we use the {\sc velocity} code of Kelson et
al. (D. Kelson, priv. comm.), which cross-correlates target and template
spectra. The errors are determined from the topology of the
cross-correlation  peak (i.e. the width of a Gaussian profile fit to
the peak in the cross-correlation spectrum) and the noise associated
with the spectrum;
typically in the range 30--210\,km\,s$^{-1}$. We present spectra of the
target radio galaxies which have
been successfully cross-correlated and shifted to the rest-frame in
Figure~2. 
In Table~3 we list the coordinates, $z'$ magnitudes and redshifts
of successfully cross-correlated targets, with the radio galaxies
highlighted.

To calculate the velocity dispersions of each cluster, we follow the iterative
method used by Lubin, Oke \& Postman (2002) and others (e.g. Willis et al. 2005).
Initially we estimate that the cluster central redshift is that of the
target radio galaxies, and select all other galaxies with $|\Delta z| <
0.06$ in redshift space. We then  calculate the biweight mean and scale
of the velocity distribution (Beers, Flynn \& Gebhardt 1990) which
correspond to the central velocity  location, $v_c$, and dispersion,
$\sigma_v$ of the cluster. We can use this to calculate the relative radial
velocities in the cluster frame:  $\Delta v = c(z-z_c)/(1+z_c)$. The
original distribution is revised, and any galaxy that lies $>$3$\sigma_v$
away from $v_c$, {\it or} has $| \Delta v | > $ 3500\,km\,s$^{-1}$ is rejected
from the sample and the statistics are re-calculated. The final result is
achieved when no more rejection is necessary, and the velocity  dispersion
and cluster redshift can be extracted. The results are presented in
Table~4, with 1-$\sigma$ errors on the cluster redshift and 
dispersion corresponding to 1000 bootstrap resamples of the redshift
distribution.

In Figure~3 we plot the redshift distribution of those galaxies for which we
performed successful cross-correlations. We also plot the velocities of those
galaxies within $|\Delta v| < 2000$\,km\,s$^{-1}$ of the cluster central
redshift, overlaid with standard Gaussian profiles. JEG\,3 is the largest
sample of cluster galaxies (in part due to the large number of galaxies
targeted in this field). JEG\,1 and JEG\,4 (at similar redshifts) have
relatively few members after the process of rejection described above.
JEG\,2 has a high velocity dispersion, indicating a non-relaxed system,
despite the fairly prominent red-sequence in this region. In part, the
high velocity dispersion might be attributed to an outlying galaxy in the velocity
distribution, but as we discuss in \S3.3, JEG\,2 might be part of a
larger, more complicated stucture than a simple group.

\begin{table*}
\caption{Spectroscopic targets with total $z'$ band magnitudes and cross-correlated
redshifts. Redshift uncertainties are derived from the topology of the
cross-correlation peak. The radio galaxy is highlighted in bold type in each case. }
\begin{tabular}{lccccc}
\hline
I.D. & RA & Dec. & $z'$ & Redshift & Member$^\star$ \cr
& (J2000.0) & (J2000.0) & (mag) \cr
\hline
\multicolumn{5}{c}{\it JEG\,1}\cr
{\bf SXDF-iE-085449} & 02:19:45.252 & $-$04:53:33.08 & 18.27 &
0.3333$^{+0.0001}_{-0.0001}$\ & $\bullet$ \cr
SXDF-iE-079345 & 02:19:39.461 & $-$04:53:18.56 & 18.83 &
0.3029$^{+0.0002}_{-0.0004}$\ & $\circ$ \cr
SXDF-iE-073826 & 02:19:36.864 & $-$04:53:29.98 & 18.01 &
0.3038$^{+0.0001}_{-0.0001}$\ & $\circ$\cr
SXDF-iE-094234 & 02:19:48.300 & $-$04:53:23.65 & 19.56 &
0.3297$^{+0.0001}_{-0.0001}$\ & $\bullet$\cr
SXDF-iE-089522 & 02:19:46.118 & $-$04:51:22.25 & 19.36 &
0.3302$^{+0.0002}_{-0.0001}$\ & $\bullet$\cr
SXDF-iE-085293 & 02:19:43.190 & $-$04:52:43.50 & 19.33 &
0.3336$^{+0.0001}_{-0.0001}$\ & $\bullet$\cr
SXDF-iE-067505 & 02:19:33.689 & $-$04:51:45.47 & 18.35 &
0.3059$^{+0.0001}_{-0.0001}$\ & $\circ$\cr
SXDF-iE-080550 & 02:19:41.508 & $-$04:52:30.60 & 17.61 &
0.3329$^{+0.0001}_{-0.0001}$\ & $\bullet$\cr
\multicolumn{5}{c}{\it JEG\,2}\cr
{\bf SXDF-iS-087629} & 02:18:23.532 & $-$05:25:00.65 & 19.17 &
0.6495$^{+0.0001}_{-0.0002}$\ & $\bullet$\cr
SXDF-iS-105254 & 02:18:19.109 & $-$05:23:06.32 & 20.02 &
0.6515$^{+0.0002}_{-0.0003}$\ & $\bullet$\cr
SXDF-iS-100288 & 02:18:19.066 & $-$05:23:45.74 & 20.19 &
0.6812$^{+0.0004}_{-0.0005}$ & $\circ$\cr
SXDF-iS-111951 & 02:18:20.074 & $-$05:22:14.13 & 21.49 &
0.6748$^{+0.0003}_{-0.0005}$ & $\circ$\cr
SXDF-iS-108679 & 02:18:20.374 & $-$05:22:36.40 & 20.64 &
0.6566$^{+0.0002}_{-0.0004}$ & $\bullet$\cr
SXDF-iS-087515 & 02:18:23.777 & $-$05:25:49.89 & 21.73 &
0.6459$^{+0.0002}_{-0.0002}$\ & $\bullet$\cr
SXDF-iS-090262 & 02:18:26.772 & $-$05:25:14.35 & 19.66 &
0.6502$^{+0.0001}_{-0.0002}$\ & $\bullet$\cr
SXDF-iS-070840 & 02:18:27.202 & $-$05:27:36.30 & 19.70 &
0.6488$^{+0.0002}_{-0.0003}$\ & $\bullet$\cr
SXDF-iS-102854 & 02:18:29.954 & $-$05:23:30.49 & 20.73 &
0.6456$^{+0.0005}_{-0.0005}$\ & $\bullet$\cr
\multicolumn{5}{c}{\it JEG\,3}\cr
{\bf SXDF-iS-170569} & 02:17:37.193 & $-$05:13:29.61 & 19.30 &
0.6472$^{+0.0004}_{-0.0004}$\  & $\bullet\dagger$ \cr
SXDF-iS-157505 & 02:17:24.230 & $-$05:15:19.53 & 19.76 &
0.6506$^{+0.0001}_{-0.0002}$\  & $\bullet$\cr
SXDF-iS-173222 & 02:17:26.530 & $-$05:13:45.93 & 21.01 &
0.6516$^{+0.0001}_{-0.0001}$\  & $\bullet$\cr
SXDF-iS-182658 & 02:17:27.502 & $-$05:11:45.34 & 21.44 &
0.6484$^{+0.0001}_{-0.0001}$\  & $\bullet$\cr
SXDF-iS-168040 & 02:17:29.052 & $-$05:12:59.61 & 19.89 &
0.6494$^{+0.0001}_{-0.0001}$\  & $\bullet$\cr
SXDF-iS-154410 & 02:17:33.286 & $-$05:15:51.00 & 19.46 &
0.6027$^{+0.0001}_{-0.0001}$\ & $\circ$\cr
SXDF-iS-184999 & 02:17:31.322 & $-$05:12:17.11 & 20.85 &
0.6010$^{+0.0001}_{-0.0001}$\ & $\circ$\cr
SXDF-iS-172233 & 02:17:35.309 & $-$05:13:30.71 & 19.76 &
0.6495$^{+0.0003}_{-0.0003}$\ & $\bullet$ \cr
SXDF-iS-172341 & 02:17:43.358 & $-$05:13:30.69 & 20.78 &
0.6470$^{+0.0001}_{-0.0001}$\  & $\bullet$\cr
SXDF-iS-172053 & 02:17:34.202 & $-$05:13:39.44 & 19.36 &
0.4464$^{+0.0002}_{-0.0002}$\ & $\circ$\cr
SXDF-iS-181772 & 02:17:25.450 & $-$05:11:54.43 & 19.81 &
0.6274$^{+0.0002}_{-0.0002}$\ & $\circ$\cr
SXDF-iS-169690 & 02:17:26.858 & $-$05:13:16.88 & 21.17 &
0.6481$^{+0.0001}_{-0.0002}$\ & $\bullet$ \cr
SXDF-iS-166516 & 02:17:31.070 & $-$05:12:37.74 & 21.17 &
0.6522$^{+0.0001}_{-0.0001}$\ & $\bullet$ \cr
SXDF-iS-167467 & 02:17:32.566 & $-$05:12:59.88 & 19.85 &
0.6433$^{+0.0001}_{-0.0001}$\  & $\bullet$\cr
SXDF-iS-178431 & 02:17:35.803 & $-$05:14:20.76 & 20.82 &
0.6475$^{+0.0002}_{-0.0002}$\  & $\bullet$\cr
SXDF-iS-163351 & 02:17:37.512 & $-$05:14:31.25 & 20.15 &
0.6455$^{+0.0002}_{-0.0002}$\  & $\bullet$\cr
SXDF-iS-172154 & 02:17:46.733 & $-$05:13:35.17 & 21.33 &
0.6456$^{+0.0007}_{-0.0005}$\  & $\bullet$\cr
SXDF-iS-131866 & 02:17:31.500 & $-$05:19:26.34 & 23.43 &
0.4943$^{+0.0002}_{-0.0002}$\ & $\circ$\cr
\multicolumn{5}{c}{\it JEG\,4}\cr
{\bf SXDF-iS-037426} & 02:18:42.067 & $-$05:32:51.03 & 18.76 &
0.3832$^{+0.0001}_{-0.0001}$ & $\bullet$\cr
SXDF-iS-029867 & 02:18:32.484 & $-$05:34:21.33 & 20.46 &
0.4238$^{+0.0002}_{-0.0002}$\ & $\circ$\cr
SXDF-iS-049475 & 02:18:34.486 & $-$05:31:28.97 & 20.51 &
0.4275$^{+0.0002}_{-0.0001}$\ & $\circ$\cr
SXDF-iS-048741 & 02:18:35.635 & $-$05:31:33.23 & 20.29 &
0.3126$^{+0.0003}_{-0.0008}$\ & $\circ$\cr
SXDF-iS-035961 & 02:18:37.447 & $-$05:32:49.61 & 18.60 &
0.4576$^{+0.0001}_{-0.0001}$\ & $\circ$\cr
SXDF-iS-041384 & 02:18:38.986 & $-$05:32:38.25 & 19.46 &
0.4725$^{+0.0002}_{-0.0002}$\ & $\circ$\cr
SXDF-iS-029332 & 02:18:40.183 & $-$05:34:01.32 & 19.31 &
0.4588$^{+0.0001}_{-0.0001}$ & $\circ$ \cr
SXDF-iS-024990 & 02:18:42.938 & $-$05:34:49.71 & 18.64 &
0.3820$^{+0.0001}_{-0.0002}$\ & $\bullet$\cr
SXDF-iS-035811 & 02:18:45.775 & $-$05:32:54.73 & 18.49 &
0.3853$^{+0.0004}_{-0.0003}$\ & $\bullet$\cr
SXDF-iS-032022 & 02:18:48.029 & $-$05:34:10.07 & 20.37 &
0.3533$^{+0.0001}_{-0.0001}$\ & $\circ$\cr
SXDF-iS-046639 & 02:18:51.972 & $-$05:31:49.76 & 20.06 &
0.3107$^{+0.0002}_{-0.0002}$ & $\circ$\cr
\hline
\multicolumn{6}{l}{$\star$~~$\bullet$ (member)~~$\circ$ (non member)}\\
\multicolumn{6}{l}{$\dagger$~~Note -- serendipitous lensed source on
  slit, $z=1.847$, see Figure 7 and \S4.2}
\end{tabular}
\end{table*}

\begin{table*}
\centering
\caption{Spectroscopic, environmental and X-ray properties of target clusters. 
The cluster redshift $z_c$ and line-of-sight velocity dispersion
$\sigma_v$ are calculated using the biweight mean and scale of the
distribution (\S3.2), with 1$\sigma$ errors corresponding to 1000
bootstrap resamples of the redshift
distribution. The environmental richness statistics (see \S3.3) --
are the $N_{0.5}$ counting statistic and $B_{\rm gc}$, the amplitude of the
galaxy-cluster cross-correlation function. We also tabulate the
results of fitting thermal ({\it mekal}) plasma models to the X-ray
spectra of the clusters (see \S3.4 for details). The uncertainties
represent 90\% for a single interesting parameter, determined during
the spectral fitting with the {\sc xspec} software.}
\begin{tabular}{lcccccccccc}
\hline
 & \multicolumn{3}{c}{Redshift properties} & \multicolumn{2}{c}{Environmental properties} &
 \multicolumn{5}{c}{X-ray properties (0.3--10\,keV)}\cr
I.D. & $z_c$ & $\sigma_v$ & $f_{\rm mem}^{\rm a}$ & $N_{\rm 0.5}$ & $B_{\rm gc}$ &
$kT$ & $Z$ & $L_X$ & $\chi^2/N_{\rm dof}$ & Net
 counts\\
      &&(km\,s$^{-1})$  &   & &(Mpc$^{1.77}$) & (keV) & ($Z_\odot$) &
      ($10^{36}$\,W) \\ 
\hline
JEG\,1 & 0.333$\pm$0.009 & 643$\pm$223 & 5/8 & $24\pm5$ & $599\pm198$
& 1.59$^{+0.65}_{-0.40}$   & \ltsim0.2  & 0.62$^{+0.07}_{-0.07}$ &
57.6/64 & $477.29\pm37.68$ \\
JEG\,2 & 0.649$\pm$0.001 & 1042$\pm$394& 7/9 & $4\pm2$ &
$696\pm557$ & 1.24$^{+4.98}_{-0.84}$ & uncon.     &
0.58$^{+0.52}_{-0.22}$ & 1.9/3 & $43.16\pm12.12$ \\
JEG\,3 & 0.648$\pm$0.001  & 774$\pm$170 & 13/18 & $23\pm5$ &
$1782\pm585$ & 1.97$^{+1.00}_{-0.59}$   & \ltsim0.7  &
1.79$^{+0.29}_{-0.83}$ & 51.2/51 & $328.21\pm33.51$ \\
JEG\,4 & 0.38$\pm$0.01 & 436$\pm$89 & 3/11 & $23\pm5$ & $388\pm204$ &
1.06$^{+0.52}_{-0.27}$   & \ltsim0.8  & 0.63$^{+0.06}_{-0.11}$ &
45.2/43 & $232.77\pm30.03$ \\
\hline
\multicolumn{9}{l}{$^{\rm a}$\,fraction of members associated with the
  radio galaxy environment for successful redshifts}\\
\end{tabular}
\end{table*}

\subsection{Environmental richness estimates}

Although a rigorous spectroscopic
environmental analysis of the clusters is not feasible with the sample
sizes in this study, we can make an estimate of the richness of the
clusters by some simple counting statistics  using the photometric
information from the {\it SXDF} catalog. The $N_{0.5}$ measure is perhaps
the best tool for a first-order quantitative analysis of a cluster
environment -- it only relies on an approximate cluster redshift (in order
to calculate the angular radius of the counting aperture), and a
satisfactorily  deep photometric catalog (Hill \& Lilly 1991). It has been
shown to correlate well with the more complex $B_{\rm gg}$ statistic
(Longair \& Seldner 1979), which is a more rigorous measure of the density
of galaxies in space when spectroscopic data is lacking, but when the shape
of the luminosity function is known (see uses in e.g. Farrah et al. 2004,
Wold et al. 2000, 2001). In physical terms $B_{\rm gg}$ is the amplitude of
the galaxy-galaxy spatial correlation function: $\xi(r) = B_{\rm
  gg}r^{-\gamma}$, where $\gamma$ is  typically 1.8--2.4 (Lilje \&
Efstathiou 1988, Moore et al. 1994, Croft, Dalton \& Efstathiou 1999),
though the low end of the range is typically chosen, with $\gamma=1.77$
(e.g. Yee \& Ellingson 2003). Both these measures can be translated to the
more familiar Abell richness scale.
 
$N_{0.5}$ is calculated by choosing a target and counting the number of
galaxies within 0.5~Mpc in the magnitude range $\left<m,m+3\right>$, where
$m$ is the magnitude of the target galaxy,  which in this case we choose to
be the radio galaxies. A field correction is applied by subtracting the
expected number of galaxies in this magnitude range from a control
field (this control field was different for each target, chosen to
be centred on a random point, $>$1\,Mpc from the radio galaxy). 
Note that the $N_{0.5}$ statistic will be inaccurate
in cases where there is strong differential evolution between the radio
galaxy luminosity and companion cluster galaxies, but this can be
compensated for by choosing the second or third brightest galaxy in the
aperture and applying the same method. More importantly, it is clear
from the spectroscopic results that the $N_{0.5}$ statistic is
contaminated by line-of-sight structures not associated with the
physical environment of the radio galaxies. This is particularly
evident for JEG\,4 -- the slightly higher redshift structure suggested by
the number of $z\sim0.4$ galaxies have magnitudes and colours that not
only re-inforce the original red-sequence selection, but will be
included in the $N_{0.5}$ calculation. Thus it is important to note
that this method can only provide a very rudimentary environmental
analysis -- a more sophisticated method is required.    

For this reason, we also calculate $B_{\rm gc}$ --  where `gc' stands
for `galaxy-cluster centre'. This is a potentially useful statistic:
derived solely from photometry, it has been shown to be in excellent
agreement with its spectroscopic counterpart,  and can be used as a
predictor for global cluster properties such as the velocity
dispersion, 
virial mass and X-ray temperature (Yee \& Ellingson 2003). Thus, we
have another parameter  to compare with our spectroscopically derived
cluster velocity dispersions and X-ray observations, as well as a
quantitative description of the richness of the radio galaxies' environments. 

A full derivation of the $B_{\rm gc}$ statistic was made by Longair \&
Seldner (1979), and has been used in a practical sense by several other
authors (see Wold et al. 2000, 2001, Farrah et al. 2003, Yee \&
L\'opez-Cruz 1999),  so we will not give a detailed procedure here, but the
form of the statistic is given by:
\begin{equation}
B_{\rm gc} = \frac{\rho_g A_{\rm gc}}{\Phi(m_{\rm lim},z)I_\gamma}d_\theta^{\gamma-3}
\end{equation}
where $\rho_g$ is the surface density of field galaxies brighter than a
limit $m_{\rm lim}$, $d_\theta$ is the angular diameter distance to $z$,
$I_\gamma$ is an integration constant ($I_{\gamma=1.77}\sim3.78$)  and
$\Phi(m_{\rm lim},z)$ is the integral luminosity function to a
luminosity corresponding to $m_{\rm lim}$ at $z$. $A_{\rm gc}$ is the
amplitude  of the angular correlation function, calculated by comparing the
net excess of galaxies within 0.5\ Mpc of the target with the number of
background sources expected for an identical aperture in the field:
$A_{\rm gc} = (N_{\rm net}/N_{\rm bg})((3-\gamma)/2)\theta^{\gamma-1}$. The
conversion to $B_{\rm gc}$ effectively deprojects $A_{\rm gc}$ from the
celestial sphere into 3-D space.

Perhaps the most important factor to consider is the choice of luminosity
function, although $B_{\rm gc}$ is reasonably tolerant of incorrect
parameters (as long as they are within $\sim20$\%  of their true values). 
We estimate the shape of the luminosity function at $z=0.35$ and
$z=0.65$ by using the semi-analytic catalog output of the 
Millenium Simulation\footnote{\tt http://galaxy-catalogue.dur.ac.uk}
(Bower et al.\ 2006, Springel et al.\ 2000). At
each epoch, we fit a Schechter function to the absolute (simulated) $R$-band
magnitude counts, fixing the faint end slope $\alpha=-0.9$. The
resulting $(M_\star,\ \phi_\star)$ are ($-20.87$ mag,\ 0.0042 Mpc$^{-3}$) and
($-20.65$ mag,\ 0.0021 Mpc$^{-3}$) at $z=0.35$ and $z=0.65$ respectively.
The relevant luminosity function used in
equation 1 is evaluated from $M_\star+3$ to $M_\star-3$, where the faint end is in a
relatively flat part of the distribution, and the bright end is expected to
include most of the cluster galaxies (note that changing these limits
by $\pm$1 magnitude does not change the value of $B_{\rm gc}$ by more
than 1$\sigma$). The results are presented in Table~4.

It is important to note that JEG\,2 appears to have an environment
almost consistent with the field (although, as noted the $N_{0.5}$ statistic is
likely to suffer contamination).
This is inconsistent with the apparent red-sequence
around the radio galaxy, which would suggest a moderate overdensity.
However, we have seen that the very high velocity dispersion for this group,
and structure in the velocity distribution might affect the richness
estimates if the group is spatially spread out. JEG\,1 and JEG\,4 are
moderately rich environments (they have very similar levels of
overdensity, although note that JEG\,4 is contaminated by a slightly
higher redshift line-of-sight structure, and so its true richness is
likely to be slightly lower),consistent with rich groups. JEG\,3 is
also an overdensity, but the statistics suggest that the cluster is no
more rich than JEG\,1 or JEG\,4. Thus, the radio galaxy selection is
detecting moderately rich groups of galaxies. 

\begin{center}
\begin{figure*}
   \centering
\includegraphics[width=\textwidth]{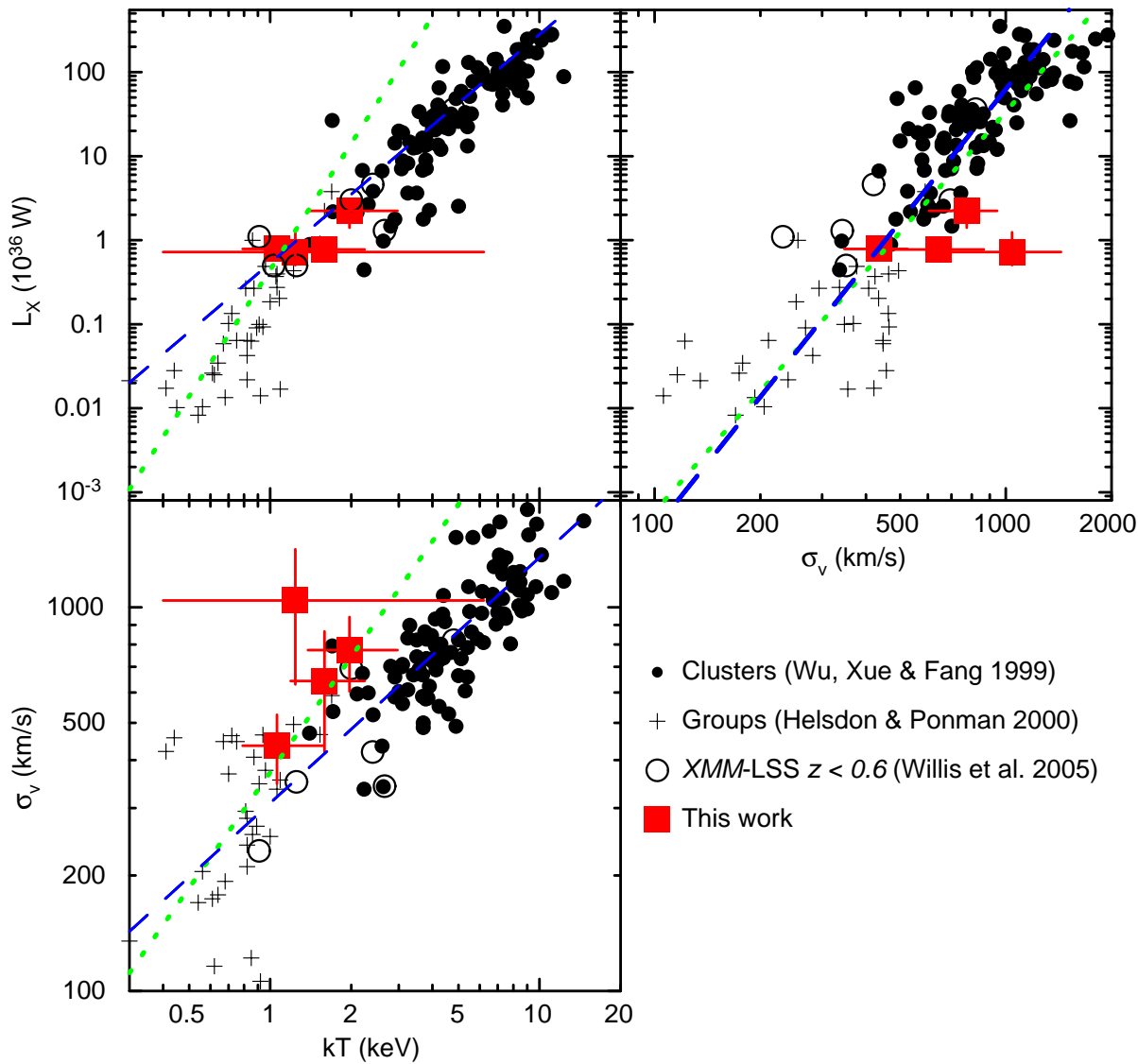}
\caption{The correlation of global cluster properties: (top left) $L_X$--$T_X$
   (bottom left) $T_X$--$\sigma_v$ (top right) $L_X$--$\sigma_v$. We compare our
   results with both rich clusters (Wu, Xue \& Fang 1999) and poor
   groups (both the `loose' and `compact' sample from Helsdon \& Ponman
   2000), as well as more recent results from the {\it XMM}-Large Scale
   Structure ({\it XMM}-LSS) survey (Willis et al.\ 2005). In all
   parameter spaces shown, the sample in this work occupy the
   intermediate space between the rich/poor sample and are similar to
   the {\it XMM}-LSS sample. The dotted lines
   are the relationships derived for the groups sample from Helsdon \&
   Ponman (2000), and the dashed lines show the equivalent relations
   derived using the rich cluster data from (Wu, Xue \& Fang 1999).
   For clarity, we do not
   plot errors on the literature points (but these are taken into
   account in the linear fits). The X-ray data from this work are corrected for
   aperture losses. To estimate this, we extrapolate a King surface brightness profile of
   the form: $S(r)  = S_0[ 1 + ({r}/{r_{\rm core}})^2]^{-3\beta +
   0.5}$. Assuming $\beta=0.5$ and $r_{\rm core} = 200$\,kpc, we find that
   the apertures used recover $\sim$80\% of the total flux,
   extrapolating to $2r_{\rm core}$. All points
   are corrected to our adopted cosmology, 
   with $H_0 = 75$\,km\,s$^{-1}$\,Mpc$^{-1}$.  }
   \label{fig:example}
\end{figure*}
\end{center}

\begin{center}
\begin{figure*}
\centering
\includegraphics[width=6.in]{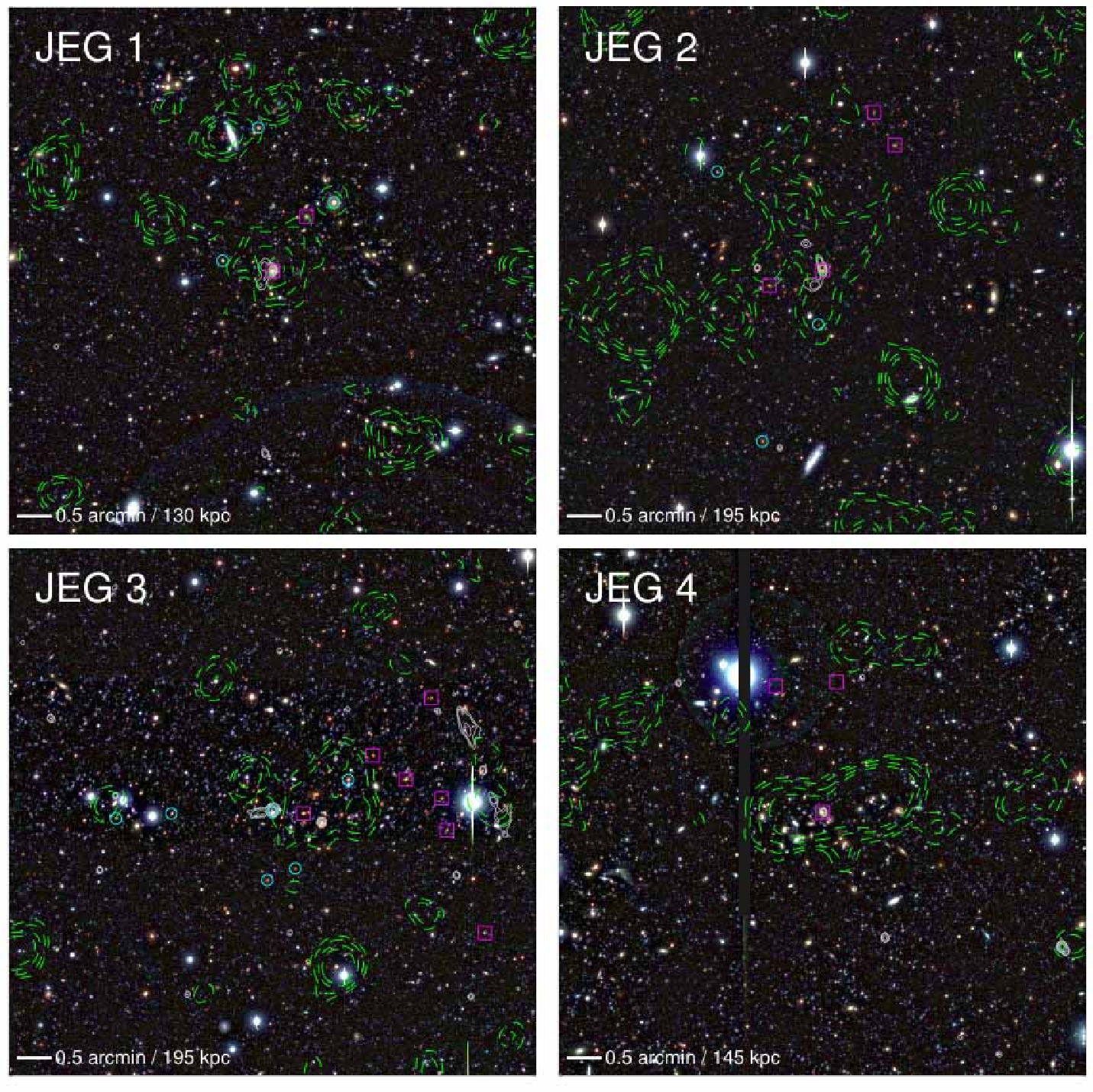}
   \caption{Here we present true colour image constructed from {\it Subaru} $BVi$-band $8'\times8'$
  images centered on the target radio galaxies JEG\,1--4 (orientated
  North up and East left). 
  In addition to the colour information, we overplot
  as contours the radio galaxy morphology from the VLA images (white solid),
  and X-ray (0.3--10\,keV) contours (green dashed). The radio contour levels
  have been chosen to emphasise the lobe morphology (note that these
  are VLA B-array maps with a beam size of $5'\times4''$ with ${\rm
    P.A.}=170^\circ$, see S06), and the X-ray
  contours are slightly smoothed with a Gaussian kernel of width $\sim$$1'$
 and at $\sim$2,3,5 \& 7$\sigma$ above the 
  background in the vicinity of the environment. 
  Finally, using the redshift data, we highlight galaxies which
  have $|\Delta v| < 2000$\,km\,s$^{-1}$ relative to the cluster redshift.
  We plot positive values of $\Delta v$ as open circles and negative values
  as open squares.}
   \label{fig:example}
\end{figure*}
\end{center}

\subsection{X-ray spectral fitting}

The X-ray spectral fitting package {\sc xspec} was used to fit X-ray
spectral models to the cluster spectra. In terms of X-rays from galaxy
clusters, it is commonly believed that the emission is predominantly
due to the hot cluster gas trapped in the potential well,
absorbed by the hydrogen column density within our own Galaxy
(i.e. little in the way of internal cluster absorption is
expected). The spectra therefore were fitted with a standard model
comprised of a {\em mekal} hot plasma emission spectrum, 
together with a {\em wabs} photo-electric absorption
model (Morrison \& McCammon 1983). Due to the small number of counts,
the hydrogen column of the absorption was fixed at the Galactic
hydrogen column density in the line of sight to the Subaru pointings:
$N_{\rm H} =2.55\times10^{20}$\,atoms cm$^{-2}$ (Dickey \& Lockman
1990). Also the redshifts were fixed at the values given in Table~4. 
The temperature and metallicity were left free to
optimize. Acceptable spectral fits were found for all the clusters
(though only a small number of counts is seen from JEG\,2), and these
results are given in Table~4. We present the best fit gas
temperature and metallicity (with respect to the Solar value),
the errors being 90\% for a single interesting parameter, and the
(0.3$-$10\,keV) absorption-corrected X-ray luminosity. 
Also tabulated are the best fit $\chi^{2}$
values and the number of degrees of freedom. For none of the clusters were
there sufficient counts to warrant the use of a more complex model. We
attempted to rebin the spectra into five counts per bin, with a fit
based on Cash statistics (i.e. not subtracting the background in order
to maintain counting statistics, Cash 1979), however this does not provide stronger
constraints on $T$ or $Z$ compared to the adopted method.
The data and best-fit models are shown in Figure~4.

\section{Discussion} 

With the exception of JEG\,2, the radio galaxies in each of our fields have
environmental properties consistent with moderately rich groups at $z\sim0.5$, in
terms of their density statistic and X-ray properties. Despite the
presence of a clear red sequence in the field of JEG\,2 (Fig.~1), a
simple $N_{0.5}$ environmental measure suggests that the
environment is consistent with the field. However, it is likely that this is
in fact a cluster of similar richness to JEG\,1, JEG\,3 and JEG\,4 given
that there are several nearby galaxies with colours and spectroscopic
redshifts close to the radio galaxy. The low-number statistics for this
system makes it difficult to make any robust comments about the
radio-galaxy environment, except that this is probably a dynamically young group.
There is some evidence for a sheet at $z\sim0.65$ (Simpson et al. in
prep) and JEG\,2 may simply lie in an overdensity within this stucture.
A large projected distribution of red galaxies around this group would in
part explain the failure of the $B_{\rm gc}$ statistic to detect an
overdensity: it could be possible that the control aperture itself is
contaminated by the larger stucture. 

Our observations of JEG\,1 and JEG\,4 also do not contain enough confirmed
spectroscopic members  at the radio galaxies' redshifts to comment on their
environments from a spectroscopic standpoint. The wide-angle radio tail of
JEG\,1 suggests rapid movement through a dense medium (S06), consistent with a
group-group merger (e.g. Jetha et al.\ 2006). 
There is insufficient spectroscopic data to confirm
this though. Furthermore, in the case of
JEG\,4, our results could be affected by a background group, although
it is not clear that this is contributing significantly to the X-ray
luminosity. This slightly more distant 
group could also be re-enforcing the strength of the red sequence (Fig.~1), 
and is a cautionary point regarding selection in this way.

\subsection{Scaling relations}

The X-ray luminosities derived by fitting thermal models to the X-ray
spectra are remarkably constant over our radio galaxy sample, with the 
environment of JEG\,3 the hottest and
intrinsically the most luminous (we note that the small number of counts seen 
from JEG\,2 results in a poorly constrained model).
In Figure 5 we compare these clusters' properties with the $L_{X}$--$T_X$ 
relation of Helsdon \& Ponman's (2000) study of galaxy groups and
low-luminosity clusters, more luminous and massive systems from Wu,
Xue \& Fang (1999), as well as more intermediate redshift ($z<0.6$)
X-ray selected groups/clusters from the {\it XMM}-Large Scale Survey
({\it XMM}-LSS; Willis et al. 2006).

Firstly, the four clusters all have high-$L_{X}$ and
high-$T$ compared to the group sample, indicating that they bridge the
gap between groups and rich clusters -- a relatively unexplored region
of parameter space. This is likely due to selection effects, thus there is
potential in investigating these types of environment using a selection
technique similar to the one outlined in this work. However, we note
that the radio selected environments are quite similar to X-ray
selected groups/clusters at similar redshifts, detected in the {\it
  XMM}-LSS (Willis et al.\ 2006). We show also the
disparity between the slopes of the $L_X$--$T_X$ relation for groups
(Helsdon \& Ponman\ 2000) and rich clusters (Wu, Xue \& Fang\ 1999),
which differ by a factor $\sim$2. This is probably a real effect -- the
steepening of the slope for groups reflects the fact that these
systems have a propsenity to exhibit deviations from the relation seen
for the more massive systems due to processes other
than cooling (Willis et al.\ 2005; Ponman, Cannon \& Navarro
1999). Note that all the empirical relations found for groups and
clusters are steeper than that expected for pure gravitational
structure formation (Kaiser 1986); this implies that feedback as a
source of energy input is important over all scales of clusters, but is
more dominant in the low mass systems.

One can also compare the cluster X-ray luminosity and velocity
dispersion with the $L_{X}$--$\sigma$ relations for these environments
(Fig.~5). Although our small numer statistics provide only loose
constraints on our environmental parameters,
all our environments appear to have slightly large velocity 
dispersions compared to the temperature of the intracluster gas, although
this is not statistically significant. Once again, the velocity dispersions
and temperatures reveal that these radio galaxies inhabit intermediate
environments between groups and clusters. A similar picture is revealed by
the $L_X$--$\sigma_v$ relationship, with all clusters having slightly large
$\sigma_v$ for their $L_X$. This is in good agreement with the observed
velocity dispersions for other radio/optically selected samples
(e.g. Rasumussen et al. 2006; Popesso et al. 2007). 

For the remainder of the discussion we concentrate on JEG\,3, which has a
spectroscopically-confirmed sample large enough to comment on the Mpc-scale
environment of the radio galaxy.

\subsection{Discussion of JEG\,3}

Optically, the radio galaxy is intriguing: it appears to be morphologically
disturbed, with a blue and red part, suggestive of a recent or ongoing
interaction triggering star-formation. In fact, closer inspection of
the spectra (and images) provide unambiguous evidence that the blue
compenent is actually a strongly lensed background AGN or starburst galaxy at $z=1.847$ -- not
associated with the group (Figure~7). This high redshift object does not affect
our results in any way, other than the fact that without the
spectroscopic confirmation, this might be classified as a merger from
the optical images alone. In terms of the radio emission, we are
confident that the majority of the emission
originates in the lensing source -- i.e. the original selection of this
target is still secure (the morphology and position of the radio
emission support this, see Fig.~6). In terms of the impact from X-ray emission
from the lensed galaxy, we also estimate that our measurements of the
X-ray luminosity of the group will not be affected. The main
bulk of the X-ray emission appears to be offset from the radio galaxy
itself (Fig.~6) and the data were cleaned for point source emission
prior to spectral fitting (\S2.2). In the case that the point source
is not cleaned, we estimate the impact on our measurement of $L_X$: if
the lensed galaxy is a starburst, then we expect the X-ray luminosity to be of order $\sim
2\times10^{34}$\,W (based on an X-ray stack of $z\sim2$ BM/BX galaxies, Reddy
\& Steidel 2004), although this will be boosted by the lens. If it is
an AGN, then we might expect the X-ray luminosity to be higher by a
factor $\sim$2 (Lehmer et al. 2005). In both cases, this has a
negligible impact on our results.  

The most obvious feature of JEG\,3's environment is the apparent spatial
offset between the radio galaxy itself and the bulk of galaxies at the group
redshift of 0.649 (Fig.~6), coincident with the dominant
source of X-ray emission.  The galaxy's radio lobe morphology is suggestive
of infall through a  dense intracluster medium towards this higher density
region. In Fig.~6  we represent the relative velocities of galaxies about the
central redshift, indicating objects with $\pm\Delta v$. Note that
there are two other radio galaxies in this group; one $\sim$$0.5'$ to
the west of the target radio galaxy, and the other (FR\,II source)
several more arcminutes further to the west. Spectroscopic observations
of these objects (Simpson et al.\ in prep) show that neither of them is
associated with the structure surrounding JEG\,3.
These radio galaxies are not
associated with the $z\sim0.65$ structure identified around the radio
galaxy in this work (see Simpson et al. in prep).

The radio galaxy  appears to be associated with a small group interacting 
with the main body of the structure, thus it is not unreasonable to associate 
the triggering of the radio galaxy with this interaction. 
This is a scenario which has been
seen in other radio galaxy environments, at both high and low redshifts, and
over a wide range of radio powers (Simpson \& Rawlings 2002). For
example, the high redshift radio galaxy TN\,J1338$-$1942 ($z=4.1$,
Venemans\ et~al.\ 2003) belongs to a `proto-cluster', but is offset
from the centre of the overdensity. 
Similarly 
at low-redshift, the FR\,II source Cygnus A ($z=0.056$) is involved in an
cluster-cluster merger (e.g. Ledlow\ et~al.\ 2005), with the radio-source
offset from the main overdensity (Owen\ et~al.\ 1997). X-ray mapping of this
cluster revealed hot (presumably) shocked gas between the two main peaks of
the X-ray emission in the two sub-cluster units (Markevitch\ et~al.\ 1999),
consistent with the model of head-on cluster merging. These well studied
radio galaxies are examples of the most luminous objects at their
redshifts, whereas the luminosities of the radio galaxies in this work are
modest in comparison. Triggering within interacting sub-cluster units seems
to be responsible for a proportion of radio-galaxies, and this appears to
have been happening at all epochs, and over a wide range of luminosities.

The triggering of the radio galaxy could have occurred via
galaxy-galaxy interactions in its immediate group, while the X-ray emission
seems to be centred on the richer group in which gas is plausibly being heated by
gravitational processes. However, there is a further complication to the
$L_X$--$T_X$ relationship, in that the outcome of the interaction between 
two sub-groups is a `boosting' of the X-ray luminosity and temperature of the
resulting cluster (Randall, Sarazin \& Ricker\ 2002). In general this
will temporarily
(i.e. for a few 100\,Myr) enhance both $L_X$ and $T_X$, before the cluster
settles down to its equilibrium value. It is not clear what stage of
interaction this environment is in. Given the relatively large offset
between the X-ray emission and the radio galaxy, we postulate that this
is in an early stage. In these environments it appears that we are
witnessing radio galaxy activity and group-group merging/interaction
at the same time. Although this may hint at radio galaxy triggering in
the cluster assembly phase, this does not imply that all clusters
containing a radio galaxy are in an unrelaxed state -- most relaxed
clusters also contain a central radio source. This is interesting in
that it appears that radio galaxy activity can be important in the
thermodynamic history of the cluster environment over a wide range of
scales (i.e. groups to massive clusters) and various stages of
development. 

Following the assumption made in equation 4 of Miley (1980), and
adopting an expansion speed of $0.03-0.15c$ (e.g. Scheuer 1995), we
estimate the jet power of JEG\,3 to be of the order
0.5--1$\times$10$^{36}$\,W, which
is comparable to the X-ray luminosity of the cluster (the expected
contribution to the X-ray luminosity from the radio galaxy itself is not
significant). 
Although the similarity in power is likely coincidental, it is interesting,
because the radio output of this low-power source appears to be
sufficient to balance the
radiative cooling of the gas. However, it is unlikely that the radio galaxy
could keep up this balance over a long period: this mechanical injection will 
end long before the main merger between the two groups stops imparting
energy to the system, given the expected timescales of radio activity
($\lesssim$$10^8$\,yr, Parma et al.\ 2002) which will likely be the dominant event in the
pre-equilibrium thermodynamic history of the cluster. Nevertheless, since
the radio-galaxy appears to have been triggered early in the life-time of
the cluster (the centres of the radio and X-ray emission are separated
by $\sim$200\,kpc, Fig.~6), it could be important for providing
significant energy input to the ICM, and therefore raising its entropy.

\begin{figure}
\centerline{\includegraphics[width=3.5in]{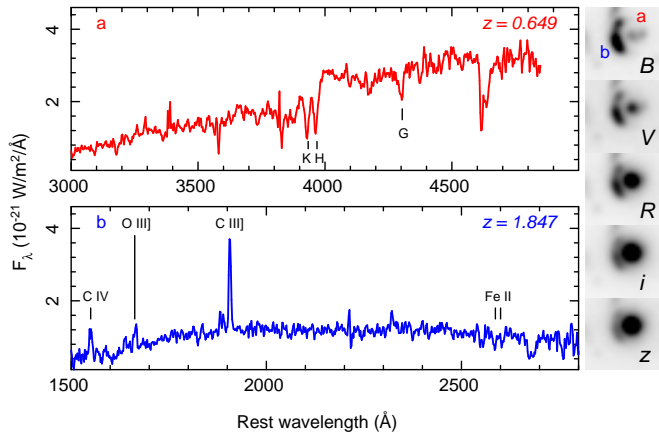}}
\caption{Spectra of components `a' and `b' of the radio galaxy in the
  field of JEG\,3 -- SXDF-iS-170569. The inset greyscale images are
  $6''\times6''$ Subaru SuprimeCam thumbnails in $BVRiz$
  bands, clearly showing the distinction between a red component `a'
  and  blue component `b'. The spectra unambiguously identify `b' with
  a galaxy at $z=1.847$, strongly lensed by the foreground component
  `a'. The emission lines at the very blue end of the
  spectrum `b' are C{\sc iv}, O{\sc iii}] and C{\sc iii}] in emission,
  securing the redshift. Fe III is seen in absorption at 2587\AA\ and
  2600\AA.}  

\end{figure}

\section{Summary}

We have presented multi-object (Low Dispersion Survey Spectrograph 2)
spectroscopy, combined with complimentary radio (VLA) and X-ray ({\it
XMM-Newton}) observations of the Mpc-scale environments of a sample of
low-power ($L_{1.4\,\rm GHz} \lesssim 10^{25}$\,W\,Hz$^{-1}$)
$z\sim0.5$ galaxies in the {\it Subaru-XMM Newton Deep Field} (Simpson
et al. 2004).

The radio galaxies targeted all appear to inhabit moderately rich groups,
with remarkably similar X-ray properties, with JEG\,3 (VLA\,0033 in Simpson
et al. 2006) being the hottest and residing in the richest environment. A
statistical interpretation of the environment of JEG\,2 (VLA\,0011 in
Simpson et al. 2006) would suggest that it does not belong to a cluster, but
this could be in part due to the spatial distribution of the galaxies
-- the other evidence that this is a relatively rich group is
compelling: there is a relatively strong red sequence and extended
X-ray emission seen at this location (suggesting a gravitationally
bound structure), and the spectroscopic data
(although dominated by small number statistics) suggests that this is indeed
a group, or part of a much larger structure, albeit one that is not
relaxed. Also, three of the radio galaxies exhibit extended lobe
morphology that suggest movement through a dense intracluster medium,
again hinting that these systems are undergoing dynamical evolution.

Our analysis concentrates on the environment of JEG\,3, which presents us
with a `snap-shot' of radio-galaxy and cluster evolution at $z\sim0.65$. 
Overall, this is the richest environment we have studied, and the
spectroscopic data reveal sub-structure: the radio galaxy belongs to a small
group of galaxies which appears to be
interacting (merging) with a larger group of galaxies. The X-ray
emission for this system is associated with the richer, but radio-quiet
group. We postulate that the radio emission has been
triggered by a galaxy-galaxy merger within its local group. Although this may
be providing some feedback to the ICM (indeed, the radio output is
capable of balancing the overall radiative cooling), the eventual group-group
interaction will boost the overall X-ray luminosity and temperature. 
This process will take
longer than the life-time of the radio-source, and so might be the dominating
event in the thermodynamic fate of the environment.  This scenario is not
dissimilar to the situation with other radio galaxies at both high redshift,
and in the local Universe. Moreover, radio-triggering within sub-cluster
units appears to be important for a proportion of the radio galaxy population
over all epochs and luminosities.

\section{acknowledgements}

We thank an anonymous referee for helpful comments that greatly
improved the clarity of this paper.
We appreciate invaluable discussions with John Stott, Tim Beers, Alastair Edge and
Mark Swinbank. The {\sc velocity} code and data reduction software were kindly
provided by Daniel Kelson. JEG, CS, SR \& AR gratefully acknowledge
support from the U.K. Science and Technology Facilities Council
(formerly the Particle Physics and Astronomy Research Council).

\setlength{\bibhang}{0.25in}

\end{document}